\documentclass[a4paper,10pt,twoside]{cpc-hepnp}
\usepackage{upgreek,fancyhdr}
\usepackage{multicol}
\usepackage{graphicx}
\usepackage{booktabs}
\usepackage{amssymb,bm,mathrsfs,bbm,amscd}
\usepackage[tbtags]{amsmath}
\usepackage{lastpage}

\begin{document}

\fancyhead[c]{\small Chinese Physics C~~~Vol. xx, No. x (201x) xxxxxx}
\fancyfoot[C]{\small 010201-\thepage}

\footnotetext[0]{Received 31 June 2015}

\title{ Joint remote preparation of an
arbitrary four-qubit \\ $|\chi\rangle$-type entangled state
\thanks{Supported by National Natural Science Foundation of China
(No. 11547023, No.11174101, No.11074088) }}

\author{%
      FU Hao $^{1)}$ (¸¶ºÆ)\email{haofu639@163.com}
 \quad CHEN Gui-Bin(³Â¹ó±ö)     \quad  LI Xiao-Wei (ÀîÏþÞ±)\\  \quad MA  Peng-Cheng (ÂíÅô³Ì) \quad
 ZHAN You-Bang (Õ²ÓÓ°î) } \maketitle

\address{ School of Physics and Electronic Electrical Engineering,
Huaiyin Normal University, Huaian 223300,  China  }

\begin{abstract}
We present two schemes for joint remote preparation of an arbitrary
four-qubit $\left\vert \chi \right\rangle $ -type entangled state
via three three- and (N+1)-qubit GHZ states as the quantum channel,
respectively. In these schemes, two senders (or N senders) share the
original state which they wish to help the receiver to remotely
prepare. To complete the JRSP schemes, several novel sets of
mutually orthogonal basis vectors are introduced. It is shown that,
only if two senders (or N senders) collaborate with each other, and
perform projective measurements under suitable measuring basis on
their own qubits respectively, the receiver can reconstruct the
original state by means of some appropriate unitary operations and
suitable C-NOT gates.  Compared with the previous schemes for the
JRSP of four-qubit $|\chi\rangle$-tpye state, the advantages of the
present schemes are that the total successful probability of JRSP
can reach 1 and the entanglement resource can be reduced.
\end{abstract}



\footnotetext[0]{\hspace*{-3mm}\raisebox{0.3ex}{$\scriptstyle\copyright$}2013
Chinese Physical Society and the Institute of High Energy Physics
of the Chinese Academy of Sciences and the Institute
of Modern Physics of the Chinese Academy of Sciences and IOP Publishing Ltd}%

\begin{multicols}{2}

\section{Introduction}

Multipartite entangled states play a fundamental role in the field
of quantum information theory and its applications. So far
multipartite entanglement has been well studied theoretically and
experimentally (\emph{e.g.} [1]-[9]). A few years ago, Verstraete
\emph{et al.} [10] have shown that the four- qubit entangled state
may be divided to nine families of states under stochastic local
operation and classical communication (SLOCC). Briegel \emph{et al}.
[11] introduced a special kind of multipartite entangled states, the
so-called cluster states. Lee \emph{et al}. [12] have constructed a
kind of four-qubit genuine entangled state $|\chi \rangle$   under
SLOCC. This state can be expressed as
\begin{eqnarray}
\left\vert \chi \right\rangle  &=&\frac{1}{2\sqrt{2}}\left(
\left\vert 0000\right\rangle -\left\vert 0011\right\rangle
-\left\vert
0101\right\rangle +\left\vert 0110\right\rangle \right.   \nonumber \\
&&\left. +\left\vert 1001\right\rangle +\left\vert 1010\right\rangle
+\left\vert 1100\right\rangle +\left\vert 1111\right\rangle \right)
_{1234}.
\end{eqnarray}
The genuine state $\left\vert \chi \right\rangle $\ has many
interesting entanglement properties. It has been shown that a new
Bell inequality is optimally violated by the genuine state
$\left\vert \chi \right\rangle $ but
not by the four-qubit GHZ state and the cluster state [13]. The state $%
\left\vert \chi \right\rangle $\ has important applications in the
quantum information field [14-19].

In the last decade, Lo [20], Pati [21], and Bennett \textit{et al.}
[22] presented a new quantum communication scheme that uses
classical communication and a previously shared entangled resource
to remotely prepare a quantum state. This communication scheme is
called remote state preparation (RSP). In RSP, Alice is assumed to
know fully the transmitted state to be prepared by Bob, so RSP is
called the teleportation [23] of a known state. Compared with the
teleportation, RSP requires less classical communication cost than
teleportation. Since then, RSP has attracted much attention, various
theoretical schemes for generalization of RSP have been proposed and
experimental implementations of RSP have been presented [24-45].
Recently, a novel aspect of RSP, called as the joint RSP (JRSP), has
been proposed [46-60]. In these schemes of the JRSP [46-60], two
senders (or $N$ senders) know partly of original state they wish to
remotely preparation, respectively. If and only if all the senders
agree to collaborate, the receiver can reconstruct the original
quantum state.  More recently, Luo and Deng [61] present a scheme
for JRSP of an arbitrary four-qubit $|\chi\rangle$-tpye entangled
state. One can note easily that, however, their JRSP scheme can not
be realized with unit success probability.

In this paper, we proposed two schemes for JRSP of an arbitrary
four-qubit $ \left\vert \chi \right\rangle $-type state with two and
$N$\ senders, respectively. To complete the JRSP schemes, several
novel sets of four-qubit measuring basis are introduced. In these
schemes, two (or $N$) senders share the original state, but each
sender only partly knows the state. It is shown that, if and only if
two (or $N$) senders agree to collaborate, the receiver can
reconstruct the original quantum state. Moreover, it is shown that,
for both present schemes, the total successful probability of JRSP
can reach 1.

\section{JRSP of an arbitrary four-qubit $\left\vert
\protect\chi \right\rangle $ entangled state with complex
coefficients}
\bigskip We present the joint remote preparation of an arbitrary four-qubit $
\left\vert \chi \right\rangle $ entangled state with complex
coefficients. In the first scheme the original state is shared by
two senders, while the prepared state is shared by N senders in the
second scheme.

\subsection{JRSP with two senders}

\bigskip Suppose that two senders Alice and Bob wish to help the receiver
Charlie remotely prepare the state

\begin{eqnarray}
\left\vert \psi \right\rangle &&=x_{0}\left\vert 0000\right\rangle
+x_{1}e^{i\delta _{1}}\left\vert 0011\right\rangle +x_{2}e^{i\delta
_{2}}\left\vert 0101\right\rangle \nonumber \\ &&+x_{3}e^{i\delta
_{3}}\left\vert 0110\right\rangle  +x_{4}e^{i\delta _{4}}\left\vert
1001\right\rangle +x_{5}e^{i\delta _{5}}\left\vert 1010\right\rangle  \nonumber \\
&&+x_{6}e^{i\delta _{6}}\left\vert 1100\right\rangle
+x_{7}e^{i\delta _{7}}\left\vert 1111\right\rangle , \label{1}
\end{eqnarray}%
where $x_{j}$ and $\delta _{j}(j=0,1,\cdots ,7)$ are real, $\delta
_{0}=0$
and $\sum_{j=0}^{7}x_{j}^{2}=1$. Assume that Alice and Bob share the state $%
\left\vert \psi \right\rangle $ and they know the state partly, that
is Alice knows $x_{j}(j=0,1,\cdots ,7)$, and Bob knows $\delta
_{j}(j=0,1,\cdots ,7)$, but Charlie does not know them at all. We
also suppose that the states shared by Alice, Bob, and Charlie as
quantum channel are four GHZ states\bigskip
\begin{eqnarray}
\left\vert \lambda_{1}\right\rangle &=&\frac{1}{\sqrt{2}}\left(
\left\vert 000\right\rangle +\left\vert 111\right\rangle \right)
_{A_{1}B_{1}C_{1}},
\nonumber \\
\left\vert \lambda_{2}\right\rangle &=&\frac{1}{\sqrt{2}}\left(
\left\vert 000\right\rangle +\left\vert 111\right\rangle \right)
_{A_{2}B_{2}C_{2}},
\nonumber \\
\left\vert \lambda_{3}\right\rangle &=&\frac{1}{\sqrt{2}}\left(
\left\vert 000\right\rangle +\left\vert 111\right\rangle \right)
_{A_{3}B_{3}C_{3}},  \label{2}
\end{eqnarray}
where the qubits $A_{i}(i=1,2,3)$ belong to Alice, qubits $%
B_{i}(i=1,2,3) $ to Bob, and qubits $C_{i}(i=1,2,3)$ to Charlie,
respectively.

In order to complete the JRSP, Alice and Bob should construct their
measuring bases. The first measuring basis chosen by Alice are two
sets of mutually orthogonal bases vectors (MOBVs) $\left\{
\left\vert \eta_{k}\right\rangle \right\} (k=0,1,\cdots ,7)$, which
are given by

\begin{eqnarray}
&&\left( \left\vert \eta_{0}\right\rangle ,\left\vert \eta
_{1}\right\rangle ,\left\vert \eta_{2}\right\rangle ,\left\vert \eta
_{3}\right\rangle ,\left\vert \eta_{4}\right\rangle ,\left\vert
\eta_{5}\right\rangle ,\left\vert \eta_{6}\right\rangle ,\left\vert
\eta_{7}\right\rangle \right) ^{T}  \nonumber \\
&=&F\left( \left\vert \xi_{0}\right\rangle ,\left\vert
\xi_{1}\right\rangle ,\left\vert \xi_{2}\right\rangle ,\left\vert
\xi_{3}\right\rangle ,\left\vert \xi_{4}\right\rangle ,\left\vert
\xi_{5}\right\rangle ,\left\vert \xi_{6}\right\rangle ,\left\vert
\xi_{7}\right\rangle \right) ^{T},  \label{3}
\end{eqnarray}%
where

\begin{equation}
F=\left(
\begin{array}{rrrrrrrr}
x_{0} & x_{1} & x_{2} & x_{3} & x_{4} & x_{5} & x_{6} & x_{7} \\
x_{1} & -x_{0} & x_{3} & -x_{2} & x_{5} & -x_{4} & x_{7} & -x_{6} \\
x_{2} & -x_{3} & -x_{0} & x_{1} & -x_{6} & x_{7} & x_{4} & -x_{5} \\
x_{3} & x_{2} & -x_{1} & -x_{0} & x_{7} & x_{6} & -x_{5} & -x_{4} \\
x_{4} & -x_{5} & x_{6} & -x_{7} & -x_{0} & x_{1} & -x_{2} & x_{3} \\
x_{5} & x_{4} & -x_{7} & -x_{6} & -x_{1} & -x_{0} & x_{3} & x_{2} \\
x_{6} & -x_{7} & -x_{4} & x_{5} & x_{2} & -x_{3} & -x_{0} & x_{1} \\
x_{7} & x_{6} & x_{5} & x_{4} & -x_{3} & -x_{2} & -x_{1} & -x_{0}%
\end{array}%
\right) ,  \label{4}
\end{equation}%
and

\begin{equation}
\begin{array}{rrrr}
\left\vert \xi_{0}\right\rangle =\left\vert 000\right\rangle , &
\left\vert \xi_{1}\right\rangle =\left\vert 001\right\rangle , &
\left\vert \xi_{2}\right\rangle =\left\vert 010\right\rangle , &
\left\vert \xi_{3}\right\rangle =\left\vert 011\right\rangle , \\
\left\vert \xi _{4}\right\rangle =\left\vert 100\right\rangle , &
\left\vert \xi_{5}\right\rangle =\left\vert 101\right\rangle , &
\left\vert \xi _{6}\right\rangle =\left\vert 110\right\rangle , &
\left\vert \xi_{7}\right\rangle =\left\vert 111\right\rangle,
\end{array}
\label{5}
\end{equation}%
\bigskip

The second measuring bases chosen by Bob are two sets of MOBVs
$\left\{ \left\vert \tau_{j}^{(k)}\right\rangle \right\} \ \
(k,j=0,1,\cdots ,7)$, which are given by

\begin{eqnarray}
\left( \left\vert \tau^{(k)}_{0}\right\rangle ,\left\vert
\tau^{(k)}_{1}\right\rangle ,\left\vert \tau^{(k)}_{2}\right\rangle
,\left\vert \tau^{(k)}_{3}\right\rangle ,\left\vert
\tau^{(k)}_{4}\right\rangle ,\left\vert \tau^{(k)}_{5}\right\rangle
,\left\vert \tau^{(k)}_{6}\right\rangle ,\left\vert
\tau^{(k)}_{7}\right\rangle \right) ^{T}  \nonumber
\end{eqnarray}
\begin{eqnarray}
=G\left( \left\vert \xi_{0}\right\rangle ,\left\vert
\xi_{1}\right\rangle ,\left\vert \xi_{2}\right\rangle ,\left\vert
\xi_{3}\right\rangle ,\left\vert \xi_{4}\right\rangle ,\left\vert
\xi_{5}\right\rangle ,\left\vert \xi_{6}\right\rangle ,\left\vert
\xi_{7}\right\rangle \right) ^{T},
\end{eqnarray}
\bigskip where $k=0,1,\cdots ,7$ and $G^{(k)}$ is a $8\times 8$
matrix,
\begin{eqnarray}
G^{(0)} &=&G(1,r_{1},r_{2},r_{3},r_{4},r_{5},r_{6},r_{7}),  \nonumber \\
G^{(1)} &=&G(r_{1},1,r_{3},r_{2},r_{5},r_{4},r_{7},r_{6}),  \nonumber \\
G^{(2)} &=&G(r_{2},r_{3},1,r_{1},r_{6},r_{7},r_{4},r_{5}),  \nonumber \\
G^{(3)} &=&G(r_{3},r_{2},r_{1},1,r_{7},r_{6},r_{5},r_{4}),  \nonumber \\
G^{(4)} &=&G(r_{4},r_{5},r_{6},r_{7},1,r_{1},r_{2},r_{3}),  \nonumber \\
G^{(5)} &=&G(r_{5},r_{4},r_{7},r_{6},r_{1},1,r_{3},r_{2}),  \nonumber \\
G^{(6)} &=&G(r_{6},r_{7},r_{4},r_{5},r_{2},r_{3},1,r_{1}),  \nonumber \\
G^{(7)} &=&G(r_{7},r_{6},r_{5},r_{4},r_{3},r_{2},r_{1},1), \label{7}
\end{eqnarray}%
where $r_{j}=e^{-i\delta _{j}}(j=1,2,\cdots ,7),$ and $
G(a_{1},a_{2},a_{3},a_{4},a_{5},a_{6},a_{7},a_{8})$ in Eq. (8) is
given by\bigskip

\begin{eqnarray}
G\left( a_{1},a_{2},a_{3},a_{4},a_{5},a_{6},a_{7},a_{8}\right) =
\nonumber
\\  \left(\begin{array}{rrrrrrrr}
a_{1} & a_{2} & a_{3} & a_{4} & a_{5} & a_{6} & a_{7} & a_{8} \\
a_{1} & -a_{2} & a_{3} & -a_{4} & a_{5} & -a_{6} & a_{7} & -a_{8} \\
a_{1} & -a_{2} & -a_{3} & a_{4} & -a_{5} & a_{6} & a_{7} & -a_{8} \\
a_{1} & a_{2} & -a_{3} & -a_{4} & a_{5} & a_{6} & -a_{7} & -a_{8} \\
a_{1} & -a_{2} & a_{3} & -a_{4} & -a_{5} & a_{6} & -a_{7} & a_{8} \\
a_{1} & a_{2} & -a_{3} & -a_{4} & -a_{5} & -a_{6} & a_{7} & a_{8} \\
a_{1} & -a_{2} & -a_{3} & a_{4} & a_{5} & -a_{6} & -a_{7} & a_{8} \\
a_{1} & a_{2} & a_{3} & a_{4} & -a_{5} & -a_{6} & -a_{7} & -a_{8}%
\end{array}%
\right).  \label{8}
\end{eqnarray}

Now let Alice first perform three-qubit projective measurement on the qubits $%
A_{1}$, $A_{2}$  and $A_{3}$ by using the basis $\left\{ \left\vert
\eta_{k}\right\rangle \right\} (k=0,1,\cdots ,7)$ and publicly
announces her measurement outcome. Next, in accordwith Alice's
result of measurement, Bob should choose one of the measuring bases
$\left\{ \left\vert \tau_{j}^{(k)}\right\rangle \right\}
(k,j=0,1,\cdots ,7)$ to measure his qubits $B_{1}$, $B_{2}$ and
$B_{3}$. After the measurement, Bob informs Charlie of his result of
measurement by the classical channel. According to Alice's and Bob's
results, Charlie can reconstruct the original state $\left\vert \psi
\right\rangle $ by C-NOT gates and suitable unitary operation. For
example, without loss of generality, assume Alice's measurement
outcome is $\left\vert \eta _{1}\right\rangle _{A_{1}A_{2}A_{3}}$,
Bob should choose measuring basis $\left\{ \left\vert \tau
_{j}^{1}\right\rangle \right\} \left( j=0,1,\cdots ,7\right) $ to
measure his qubits $B_{1}$, $B_{2}$  and $B_{3}$, and then inform
Charlie of his measurement result by classical channel.
\bigskip Assume result of Bob's measurement is $\left\vert \tau
_3^{(1)}\right\rangle _{B_{1}B_{2}B_{3}}$, the qubits $C_{1}$,
$C_{2}$ and $C_{3}$ will be collapsed into the state
\begin{eqnarray}
|u\rangle=\frac{1}{2\sqrt{2}}(x_1e^{i\delta_1}|000\rangle+x_0|001\rangle-x_3e^{i\delta_3}|010\rangle\nonumber \\
-x_2e^{i\delta_2}|011\rangle-x_5e^{i\delta_5}|100\rangle-x_4e^{i\delta_4}|101\rangle\nonumber \\
+x_7e^{i\delta_7}|110\rangle+x_6e^{i\delta_6}|111\rangle)_{C_1C_2C_3}.
\end{eqnarray}
According to Alice's and Bob's public announcements, Charlie can
perform the local unitary operation
$(\sigma_z)_{C_1}\otimes(\sigma_z)_{C_2}\otimes(\sigma_x)_{C_3}$ on
his  qubits $C_1, C_2$ and $C_3$,   the   state $|u\rangle$ can
evolve as

\begin{eqnarray}
|u'\rangle=\frac{1}{2\sqrt{2}}(x_0|000\rangle+x_1e^{i\delta_1}|001\rangle+x_2e^{i\delta_2}|010\rangle\nonumber \\
+x_3e^{i\delta_3}|011\rangle+x_4e^{i\delta_4}|100\rangle+x_5e^{i\delta_5}|101\rangle\nonumber \\
+x_6e^{i\delta_6}|110\rangle+x_7e^{i\delta_7}|111\rangle)_{C_1C_2C_3}.
\end{eqnarray}
Next, Charlie introduces an auxiliary qubit $C_4$ with initial state
of $|0\rangle$, and the state (11) will be described as
\begin{eqnarray}
|u''\rangle=\frac{1}{2\sqrt{2}}(x_0|000\rangle+x_1e^{i\delta_1}|001\rangle+x_2e^{i\delta_2}|010\rangle\nonumber \\
+x_3e^{i\delta_3}|011\rangle+x_4e^{i\delta_4}|100\rangle+x_5e^{i\delta_5}|101\rangle\nonumber \\
+x_6e^{i\delta_6}|110\rangle+x_7e^{i\delta_7}|111\rangle)_{C_1C_2C_3}\otimes|0\rangle_{C_4}.
\end{eqnarray}
Then Charlie employs in turn three C-NOT gates $C_{C_3-C_4}$,
$C_{C_2-C_4}$ and  $C_{C_1-C_4}$ on the qubits $C_1, C_2, C_3$ and
$C_4$, where  $C_{i-j}$ denotes that $i$ as control qubit and $j$ as
target  one. After that, the state $|u''\rangle$ (see Eq.(12)) can
be transformed into the desired state $|\psi\rangle$ and the JRSP
succeeds in this situation. If Alice's measurement outcomes are the
other 7 cases in the basis $\{|\eta_k\rangle\}
(k=0,1,\cdot\cdot\cdot,7)$ , Bob should choose appropriate measuring
bases $\{|\tau_j^{(k)}\rangle\} (k,j=0,1,\cdot\cdot\cdot,7)$ to
measure his qubits  $B_1, B_2$ and $B_3$. The corresponding relation
of Alice's measurement result $|\eta_k\rangle_{A_1A_2A_3}$ and the
measuring basis $\{|\tau_j^{(k)}\rangle\}$ performed by Bob can be
described as
$|\eta_k\rangle_{A_1A_2A_3}\rightarrow\{|\tau_j^{(k)}\rangle\}(k,j=0,1,\cdot\cdot\cdot,7)$.
Similar to above approach, after Alice's and Bob's measurements,
Charlie can reconstruct the original state $|\psi\rangle$ by
appropriate unitary operation at his side. It is easily found that,
for all the 64 measurement results of Alice and Bob, the receiver
Charlie can reconstruct the original state $\left\vert \psi
\right\rangle $ by appropriate unitary operations, the success
probability of the JRSP process being 1. The required classical
communication cost is 6 bits in the scheme.\bigskip

\subsection{JRSP with N senders}

\bigskip The scheme in section 2.1 can be generalized to the case of N senders.
Suppose that Alice and Bob$_{1}$, Bob$_{2}$, . . . , Bob$_{N-1}$
wish to
help the receiver Charlie remotely prepare an arbitrary four-qubit $%
\left\vert \chi \right\rangle $\ entangled state

\begin{eqnarray}
\left\vert \phi \right\rangle =x_{0}\left\vert 0000\right\rangle
+x_{1}e^{i\varphi_{1}}\left\vert 0011\right\rangle +x_{2}e^{i\varphi
_{2}}\left\vert 0101\right\rangle \nonumber \\
+x_{3}e^{i\varphi_{3}}\left\vert 0110\right\rangle
+x_{4}e^{i\varphi_{4}}\left\vert 1001\right\rangle
+x_{5}e^{i\varphi_{5}}\left\vert 1010\right\rangle \nonumber
\\+x_{6}e^{i\varphi _{6}}\left\vert 1100\right\rangle
+x_{7}e^{i\varphi_{7}}\left\vert 1111\right\rangle ,
\end{eqnarray}
where $x_{j}$ and $\varphi_{j}(j=0,1,\cdots ,7)$ are real, $\varphi
_{0}=0$ and $\sum_{j=0}^{7}x_{j}^{2}=1$.  Assume that the N senders
know the state $\left\vert \phi \right\rangle $ partly, i.e. Alice
knows $ x_{j}(j=0,1,\cdots ,7)$, Bob$_{1}$ knows $\varphi
_{j}^{(1)}$, Bob$_{2}$ knows $\varphi_{j}^{(2)}$, $\cdots $,
Bob$_{N-1}$ knows $\varphi_{j}^{(N-1)}$, where $\varphi_{j}=\varphi
_{j}^{(1)}+\varphi_{j}^{(2)}+\cdots +\varphi
_{j}^{(N-1)}(j=0,1,\cdots ,7)$, but Charlie does not know them at
all. We also suppose that the N sender and receiver Charlie share three $(N+1)$%
-qubit GHZ states as the quantum channel, which are given by
\begin{eqnarray}
\left\vert \varepsilon_{1}\right\rangle &=&\frac{1}{\sqrt{2}}\left(
\left\vert 0\right\rangle^{\otimes (N+1)}+\left\vert
1\right\rangle^{ \otimes (N+1)}\right)
_{A^{(1)}B_{1}^{(1)}B_{2}^{(1)}\cdots B_{N-1}^{(1)}C^{(1)}},  \nonumber \\
\left\vert \varepsilon_{2}\right\rangle  &=&\frac{1}{\sqrt{2}}\left(
\left\vert 0\right\rangle^{ \otimes (N+1)}+\left\vert 1\right\rangle
^{ \otimes (N+1)}\right)
_{A^{(2)}B_{1}^{(2)}B_{2}^{(2)}\cdots B_{N-1}^{(2)}C^{(2)}},  \nonumber \\
\left\vert \varepsilon_{3}\right\rangle  &=&\frac{1}{\sqrt{2}}\left(
\left\vert 0\right\rangle ^{ \otimes (N+1)}+\left\vert
1\right\rangle ^{ \otimes (N+1)}\right)
_{A^{(3)}B_{1}^{(3)}B_{2}^{(3)}\cdots B_{N-1}^{(3)}C^{(3)}},\nonumber \\
\end{eqnarray}
where qubits $A^{(1)}$, $A^{(2)}$ and $A^{(3)}$  belong to Alice,
qubits $B_{1}^{(1)}$, $B_{1}^{(2)}$   and $B_{1}^{(3)}$
  belong to Bob$_{1}$, $\cdots $, qubits $B_{N-1}^{(1)}$, $B_{N-1}^{(2)}$ and $
B_{N-1}^{(3)}$  belong to Bob$_{N-1}$, and $C^{(1)}$, $%
C^{(2)}$ and $C^{(3)}$    belong to Charlie, respectively. As in the
above scheme, the N senders must construct their own measurement
basis. The first measuring basis chosen by Alice is still in Eqs.
(4) and (5), and the measuring bases chosen by Bob$_{1}$, Bob$_{2}$,
$\cdots $, Bob$_{N-1}$ are $ 8(N-1)$ sets of MOBVs $\left\{
\left\vert \tau _{jl}^{(k)}\right\rangle \right\} (k=0,1,\cdots,7;
j=0,1,\cdots ,7; l=1,2,\cdots ,N-1)$, which are given\bigskip\ by
\begin{eqnarray}
\left( \left\vert \tau_{0l}^{\left( k\right) }\right\rangle ,
\left\vert \tau_{1l}^{\left( k\right) }\right\rangle ,\left\vert
\tau_{2l}^{\left( k\right) }\right\rangle
,\left\vert\tau_{3l}^{\left( k\right) }\right\rangle ,\left\vert
\tau_{4l}^{\left( k\right) }\right\rangle ,\left\vert
\tau_{5l}^{\left( k\right) }\right\rangle ,\left\vert
\tau_{6l}^{\left( k\right) }\right\rangle ,\left\vert
\tau_{7l}^{\left( k\right) }\right\rangle \right) ^{T}  \nonumber
\end{eqnarray}
\begin{eqnarray}
=H_{l}^{(k)}\left( \left\vert \xi _{0}\right\rangle ,\left\vert
\xi_{1}\right\rangle ,\left\vert \xi_{2}\right\rangle ,\left\vert
\xi_{3}\right\rangle ,\left\vert \xi _{4}\right\rangle ,\left\vert
\xi_{5}\right\rangle ,\left\vert \xi_{6}\right\rangle ,\left\vert
\xi_{7}\right\rangle \right) ^{T},
\end{eqnarray}
where $\left\vert \xi_{m}\right\rangle (m=0,1,\cdots ,7)$ are given
by Eq. (6),  $ l=1,2,\cdots ,N-1$, and

\begin{eqnarray}
H_{l}^{(0)}
&=&H(1,r_{1l},r_{2l},r_{3l},r_{4l},r_{5l},r_{6l},r_{7l}),
\nonumber \\
H_{l}^{(1)}
&=&H(r_{1l},1,r_{3l},r_{2l},r_{5l},r_{4l},r_{7l},r_{6l}),
\nonumber \\
H_{l}^{(2)}
&=&H(r_{2l},r_{3l},1,r_{1l},r_{6l},r_{7l},r_{4l},r_{5l}),
\nonumber \\
H_{l}^{(3)}
&=&H(r_{3l},r_{2l},r_{1l},1,r_{7l},r_{6l},r_{5l},r_{4l}),
\nonumber \\
H_{l}^{(4)}
&=&H(r_{4l},r_{5l},r_{6l},r_{7l},1,r_{1l},r_{2l},r_{3l}),
\nonumber \\
H_{l}^{(5)}
&=&H(r_{5l},r_{4l},r_{7l},r_{6l},r_{1l},1,r_{3l},r_{2l}),
\nonumber \\
H_{l}^{(6)}
&=&H(r_{6l},r_{7l},r_{4l},r_{5l},r_{2l},r_{3l},1,r_{1l}),
\nonumber \\
H_{l}^{(7)}
&=&H(r_{7l},r_{6l},r_{5l},r_{4l},r_{3l},r_{2l},r_{1l},1), \label{14}
\end{eqnarray}%
where $r_{jl}=e^{-i\varphi_j^{(l)}}$
$(j=0,1,\cdot\cdot\cdot,7,l=1,2,\cdot\cdot\cdot,N-1)$ , and
$H(a_1,a_2,a_3,a_4,a_5,a_6,a_7,a_8)$ is also a $8\times8$ matrix
which similar to Eq.(9).

 Alice first performs the three-qubit projective measurement on her
qubits $A^{(1)}, A^{(2)}$ and $A^{(3)}$ under the basis
$\{|\eta_k\rangle\}$[see Eqs.(4) and (5)] and publicly announces her
result of measurement. According to   Alice's outcome, Bob$_1$ ,
Bob$_2$ ,$\cdot\cdot\cdot$, Bob$_{N-1}$ should choose suitable
measuring basis in the MOBVs
$\{|\tau_{jl}^{(k)}\rangle\}(k,j=0,1,\cdot\cdot\cdot,7,l=1,2,\cdot\cdot\cdot,N-1)
$ to measure their own qubits  $(B_1^{(1)},B_1^{(2)},B_1^{(3)})$ ,
$(B_2^{(1)},B_2^{(2)},B_2^{(3)})$, $ \cdot\cdot\cdot$,
$(B_{N-1}^{(1)},B_{N-1}^{(2)},B_{N-1}^{(3)})$,  and then inform
Charlie of their measurement results, respectively.  In accord with
the announcement of N senders, the receiver Charlie can reconstruct
the original state $|\phi\rangle$ by using appropriate unitary
operation and C-NOT gates. For example, without loss of generality,
suppose that Alice's measurement result is
$|\eta_0\rangle_{A_1A_2A_3}$, then Bob$_1$ , Bob$_2$
,$\cdot\cdot\cdot$, Bob$_{N-1}$ should choose suitable measuring
bases $\{|\tau_{j1}^{(0)}\rangle\} $, $\{|\tau_{j2}^{(0)}\rangle\},
\cdot\cdot\cdot $, $\{|\tau_{j(N-1)}^{(0)}\rangle\} $ [see Eqs.(15)
and (16)] to measure their own qubits, respectively. Assume that the
Bob$_1$'s measurement result is only
$|\tau^{(0)}_{11}\rangle_{B_1^{(1)}B_1^{(2)}B_1^{(3)}}$ while all
other senders' results are
$|\tau^{(0)}_{0m}\rangle_{B_m^{(1)}B_m^{(2)}B_m^{(3)}}(m=2,3,\cdot\cdot\cdot,N-1)$,
respectively, the qubits $C^{(1)},C^{(2)}$ and $C^{(3)}$ will be
collapsed into the state
\begin{eqnarray}
|v\rangle=\frac{1}{\sqrt{2}}(x_0|000\rangle-x_1e^{i\varphi_1}|001\rangle+x_2e^{i\varphi_2}|010\rangle\nonumber \\
-x_3e^{i\varphi_3}|011\rangle+x_4e^{i\varphi_4}|100\rangle-x_5e^{i\varphi_5}|101\rangle\nonumber \\
+x_6e^{i\varphi_6}|110\rangle-x_7e^{i\varphi_7}|111\rangle)_{C^{(1)}C^{(2)}C^{(3)}}.
\end{eqnarray}
According to the results of N senders,  Charlie can perform the
unitary operation
$(I)_{C^{(1)}}\otimes(I)_{C^{(2)}}\otimes(\sigma_z)_{C^{(3)}}$ on
the qubits  $C^{(1)},C^{(2)}$ and $C^{(3)}$,    the   state
$|v\rangle$ [see Eq.(17)] will evolve as
\begin{eqnarray}
|v'\rangle=\frac{1}{\sqrt{2}}(x_0|000\rangle+x_1e^{i\varphi_1}|001\rangle+x_2e^{i\varphi_2}|010\rangle\nonumber \\
+x_3e^{i\varphi_3}|011\rangle+x_4e^{i\varphi_4}|100\rangle+x_5e^{i\varphi_5}|101\rangle\nonumber \\
+x_6e^{i\varphi_6}|110\rangle+x_7e^{i\varphi_7}|111\rangle)_{C^{(1)}C^{(2)}C^{(3)}}.
\end{eqnarray}
Next, Charlie introduces an auxiliary qubit $C^{(4)}$ with initial
state of $|0\rangle$, and makes in turn three C-NOT gates
$C_{C^{(3)}-C^{(4)}}$, $C_{C^{(2)}-C^{(4)}}$ and
$C_{C^{(1)}-C^{(4)}}$ on the qubits $C^{(1)},C^{(2)}, C^{(3)}$ and
$C^{(4)}$, the desired state $|\phi\rangle$ can be recovered.

 If Alice's measurement results
are the other 7 cases in the basis $\{|\eta_k\rangle\}
(k=0,1,\cdot\cdot\cdot,7)$, Bob$_1$ , Bob$_2$ ,$\cdot\cdot\cdot$,
Bob$_{N-1}$  should choose suitable measuring bases
$\{|\tau_{jl}^{(k)}\rangle\} (k,j=0,1,\cdot\cdot\cdot,7,
l=1,2,\cdot\cdot\cdot,N-1)$ to measure their own qubits
respectively, then Charlie can recover the original state
$|\phi\rangle$ by appropriate unitary operations and suitable C-NOT
gates. Here we no longer depict them one by one. The corresponding
relation of Alice's measurement outcome
$|\eta_k\rangle_{A^{(1)}A^{(2)}A^{(3)}}$ and the measuring basis
$\{|\tau_{jl}^{(k)}\rangle\}$ performed by Bob$_1$ , Bob$_2$
,$\cdot\cdot\cdot$, Bob$_{N-1}$  can be described as

\begin{eqnarray}
|\eta_k\rangle_{A^{(1)}A^{(2)}A^{(3)}} \longrightarrow
\{|\tau_{jl}^{(k)}\rangle\},
\end{eqnarray}
where $k,j=0,1,\cdot\cdot\cdot,7, l=1,2,\cdot\cdot\cdot,N-1.$ In
this scheme, the total successful probability of the JRSP is still
1, and the required classical communication cost is $3N$ bits.

\section{Conclusion} In conclusion, we have presented two new
schemes for joint remote preparation of an arbitrary four-qubit
$|\chi \rangle$-type entangled states. In these schemes, the
coefficients of the original states to be co-prepared are all
complex. In the first scheme, two sender share an arbitrary
four-qubit   $|\chi \rangle$-type state, but each sender only partly
knows the state, and three three-qubit GHZ states are exploited as
the quantum channel. In order to help the receiver remotely prepare
the original state, in accord with the knowledge of the original
state which she/he known, each sender must construct her/his own
three-qubit measuring basis. Firstly, a sender performs a
three-qubit projective measurement on her qubits, then, another
sender should choose, according to the measurement result of the
first sender, an appropriate three-qubit measuring basis to measure
his qubits. After these projective measurements, the receiver can
reconstruct the original state by means of appropriate unitary
operations and suitable C-NOT gates. Next, we generalize the scheme
to N senders' case.  In the generalized scheme, the original state
is shared by the N senders and the quantum channel shared by the N
senders and the receiver are three (N+1)-qubit GHZ states. It is
shown that, only if when N senders collaborate with each other, the
receiver can remotely reconstruct the original state. To complete
the JRSP schemes, some novel sets of three-qubit mutually orthogonal
basis vectors have been introduced. After the projective
measurements by two senders (or N senders) under these bases
respectively, the original state can be recovered by the receiver.
Compared with the previous scheme [61], the advantage of our schemes
is that  the total successful probability of JRSP can reach 1.
Furthermore, in the present schemes, the quantum channel only
requires three three- or (N+1)-qubit GHZ states, respectively, that
is, the entanglement resource can be reduced. Thus, we hope that our
schemes will be helpful in the deeper understanding of the process
of RSP, and may be useful for the further studies on quantum
information science, such as quantum secret sharing and quantum
network communication.

\end{multicols}

\vspace{15mm}

\vspace{-1mm}
\centerline{\rule{80mm}{0.1pt}}
\vspace{2mm}

\begin{multicols}{2}

\end{multicols}

\clearpage

\end{document}